# Environmental Co-design: Fish-Blade Collision Model for Hydrokinetic Turbines

Eshwanth Asok [a], Ruo-Qian Wang [a]

[a] *Department of Civil and Environmental Engineering, Rutgers University, New Brunswick, NJ*

**Abstract**

A major challenge in the deployment of hydrokinetic turbines in aquatic environments is the risk of fish collisions. Traditional fish collision models often oversimplify this risk by neglecting critical factors, such as the thickness of the turbine and accessory structures. Additionally, variations in fish size and species are frequently overlooked. This study addresses these gaps by developing a swimming mechanics-based fish-blade collision model. Using a Lagrangian particle tracking approach, we simulate fish movements and evaluate collision risks with a representative hydrokinetic turbine, both with and without ducts. The model is applied to the velocity field at Baton Rouge, Louisiana, allowing for the assessment of collision risks across different fish species. The results offer valuable insights for turbine siting, optimization of turbine placement, and evaluation of protective designs to reduce environmental impacts in complex flow environments.

## 1. Introduction

The global energy demand is increasing due to population growth, industrial developments, and rising greenhouse gas (GHG) emissions. The EIA predicts a 50% spike in global energy demand by 2050 [1], leading to a spike in GHG emissions. Renewable energy deployments that have helped reduce $CO_2$ emissions are still struggling to meet the global energy demand that is growing faster. There is a pressing need to accelerate the transition to a renewable energy-dominant world that significantly reduces the GHGs in the next decade [2]. Hydropower, a dominant renewable energy technology using high dams to create potential head to generate power significantly impacts the ecosystem and often involves migrating communities and heritage sites [3]. As a sustainable alternative, hydrokinetic turbines, utilizing velocity head from water currents to generate power in rivers, estuaries, and coastal waters [4], [5], [6], [7] provide a cost-effective, environmentally friendly solution to hydropower technology. Recent research [8] suggests that this zero-potential-head technology project could supplement over 10% of the growing energy demand.

Environmental Impact Assessments that govern the permitting process [9] for this riverine technology typically require the evaluation of the potential impact on fish populations, and the collision risk assessment is the key to establishing such estimates. Regulatory agencies need robust data and risk models to determine if the proposed hydrokinetic projects pose an unacceptable threat to vulnerable fish species [10] before granting the permit. Moreover, understanding these collision risk assessments can guide better turbine methods to mitigate the potential impacts. These models can also help determine the suitable location of deployment and operating parameters needed.

Hydrokinetic turbine technology usually involves one or an array of horizontal or vertical-axis turbines [11], increasing underwater river complexity [12] and attracting aquatic populations towards steep flow velocity gradients [13],[14]. This could lead to non-trivial collision risks such as physical contact between devices and organisms to aquatic organisms, diving birds, and flying animals [15]. Turbine rotor blades are aquatic organisms' most common sources of injury and mortality [16]. The severity of the strike depends on factors such as the animal's ability to navigate, water velocity, number of blades, blade movement rate, and the portion of the animal hitting [15]. Horizontal axis turbine blade velocity increases from hub to tip, and the probability of the animal surviving the collision reduces from hub to tip. However, the tip speed is limited due to the cavity effect that lowers the turbine efficiencies [17], [18].

Traditional design methods emphasize maximizing power generation efficiency, and the impact on the environment is considered secondary. To obtain the permit to deploy and develop strategies for the market, a product is usually modified to sacrifice performance to meet environmental requirements. This design-for-efficiency and compromise-for-environment strategy usually leads to sub-optimal design and suffers a risk that the environmental requirements might not be met to justify the cost-benefit balance. Therefore, we propose here a new concept of Environmental Co-design. This new strategy incorporates the environmental impact and requirements in the early design phase so that no significant performance needs to be sacrificed at the final stage of design or deployment. As the first study in this new concept, we focus on fish collision risk to demonstrate this new design strategy.

Simple fish-blade collision models have been developed in the past decades [19],[20],[21],[22], but a few key factors were overlooked, compromising practical estimate of fish collision risks. First, they mainly involve conventional hydroelectric turbines that are enclosed in turbine housings of dams and do not account for emerging class of hydrokinetic turbines, which operate in open, free flowing water. Second, the developed models oversimplify the blade to a line, neglecting the span width of the blade, and there is less angle range for the animals to strike. We only know little about the magnitude of impact forces that cause injuries to the aquatic beings [23],[24] or the swimming behavior they may use to avoid or escape the strike. Such data is widely available before it will be easy to model. In addition to the blade strike, the animals can be injured by the sudden water pressure change due to the movement of the blades [25]. Third, few studies quantified the mitigation measures to reduce the collision risk and how much the performance of the turbine will be compromised. For example, [26] suggested that fish can avoid strikes in a slowly rotating turbine, but no quantitative assessment has been performed to evaluate the reduction in performance.

Unlike hydropower turbines, hydrokinetic devices operate in high-velocity, open-flow environments where fish have greater freedom to avoid turbine blades, yet collision risk remains a critical concern. Existing models attempt to assess the risk but exhibit key limitations. Some incorporate avoidance via burst swimming behavior [27] or probabilistic frameworks [28], while others neglect avoidance due to uncertainties [29],[30]. Even the most advanced models integrating Computational Fluid Dynamics (CFD) fail to account for fish behavior, leaving significant gaps in understanding fish-turbine interactions [30]. [31] developed a probabilistic model to evaluate fish collision risks, highlighting the importance of species-specific behaviors and environmental conditions. [32] further explored these impacts, with a focus on understanding the variability of risk across different species and environments. [33] introduced an agent-based model predicting fish collisions with tidal turbines, emphasizing movement patterns as key risk determinants. However, these models are either too costly in computation and require overwhelming details of the fish and blade geometry, or too much simplified to compromise the accuracy of the model. So a simplified yet effective risk assessment model is needed so that fish dimensions using site-specific empirical data are incorporated without the complex behavioral modeling. Such a model, validated with real-world fish and river flow data, would enable a more precise evaluation of collision risks and provide insights to guide permitting and turbine deployment locations while minimizing environmental impacts.

Various hydrokinetic turbines have been designed [34]. For example, unducted (Fig 1b), and ducted (Fig 1a) horizontal turbines have been deployed for commercial applications in several places in the world. This study uses a low-cost design solution for hydrokinetic energy harvesting as an example of demonstration. Specifically, the turbine is a 5 kW, ducted 3-blade reconfigurable turbine [35], similar to that shown in Fig 1a. The aim is to evaluate the probability of fish strike for ducted turbine design, as there is no simple collision risk assessment study, particularly for this design. A modified probabilistic theoretical model based on Von Raben's kinematic blade theory [36] is developed, and mitigative measures are suggested based on the modeling results.

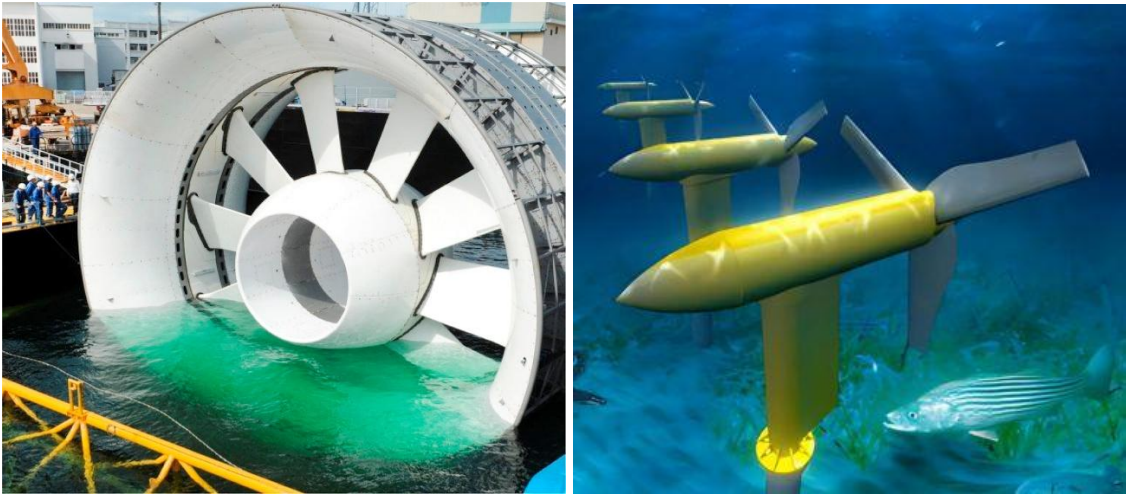

**Fig. 1.** Hydrokinetic Turbine Designs: a) Ducted Turbine by OpenHydro [37] b) Open turbine [38]

## 2. Methodology

The objective of this paper is to develop a more accurate fish-blade collision model and use the river section that the US Department of Energy targeted for hydrokinetic turbine testing [39]. The probabilistic fish-blade collision model is developed based on the kinematic blade strike model initially proposed by Von Raben [36]. He developed this method to estimate fish mortality due to turbine blade strikes in Hydropower facilities and applied it to different types of turbine designs including Francis and Kaplan turbines,

$$P_{str} = (nNL\cos(\alpha))/V_{axial} \tag{1}$$

where $L$ is the length of fish approaching the leading edge of the turbine, $\alpha$ is the angle of fish relative to the turbine rotor, $V_{axial}$ is the axial water velocity at turbine blades, $n$ is the number of blades, and $N$ is the rotations per second of the turbine. Based on this, we have developed a modified model to evaluate the collision risk based on the force balance applied particularly for the Hydrokinetic blade design.

### 2.1 Theoretical Collison risk model

As the turbines in this technology are rotating with a high blade velocity, the aquatic animals passing the waterway would have a non-trivial collision risk. A Lagrangian tracking model is developed to simulate the fish's swimming dynamics. This simplification allows the focus to remain on variable geometry and rotor interactions. Specific geometric data for fishes, including size and width, are collected based on the various species listed at this site [40]. These inputs are crucial for calibrating the model to local fish and physical characteristics.

We assume the fish can be approximated as a sphere that is neutrally buoyant and only subject to the swimming force and the drag. The drag force $F_d$ acting on fish can be represented as

$$F_d = 0.5 C_d A_f \rho U_f^2 \qquad (2)$$

where $C_d$ is the drag coefficient, $A_f$ is the cross-sectional area of fish ($A_f=V_f/L$), $U_f$ is the relative velocity of fish. Since our river flow is turbulent with a high Reynolds number, $C_d$ is considered to be 0.01.

We further assume that the swimming force by the fish $F_s$ is proportional to the fish's volume $V_f$ is given by

$$F_s = C_s \rho_f g V_f \qquad (3)$$

where $C_s$ is the swimming coefficient of fish which is assumed to be 0.015 [41], $g$ is the gravitational constant, and $V_f$ is the volume of fish, which is estimated by assuming fishes as spheres. We further consider the aspect ratio of the fish, i.e. the ratio of fish width to fish length, is 1/3.

The new version of strike probability depending on an unducted RAFT turbine blade design is:

$$P_{str} = \frac{\text{Fish passing time}}{\text{Blade sweeeping time}} = \frac{(T+L\sin(\alpha))/(U_f \sin(\alpha)+U_0)}{1/nN} \qquad (4)$$

where $L$ is the mean length of the fish approaching the turbine, $\alpha$ is the angle of the fish relative to the turbine rotor, $U_f$ is the velocity of the fish, $n$ is the number of blades, $N$ is the rotations per second of the turbine, and $U_0$ is the velocity of the river as shown in Fig 2. We used the above force equations to derive a relation for the fish velocity ($U_f$) and used it in the modified probability of strike equation. Normalization of the equation leads to the derived relation for the probability of strike given below and can be used for evaluation based on varying conditions.

$$P_{str} = M \frac{1+L_x \sin(\alpha)}{1+k\sqrt{L_x}\sin(\alpha)} \qquad (5)$$

where the normalized rotation speed $M = \frac{nNT}{U_0}$, the "collision coefficient" $k = \sqrt{\frac{2C_s \rho_f gT}{C_d \rho U_0^2}}$ and the normalized fish length $L_x = L/T$

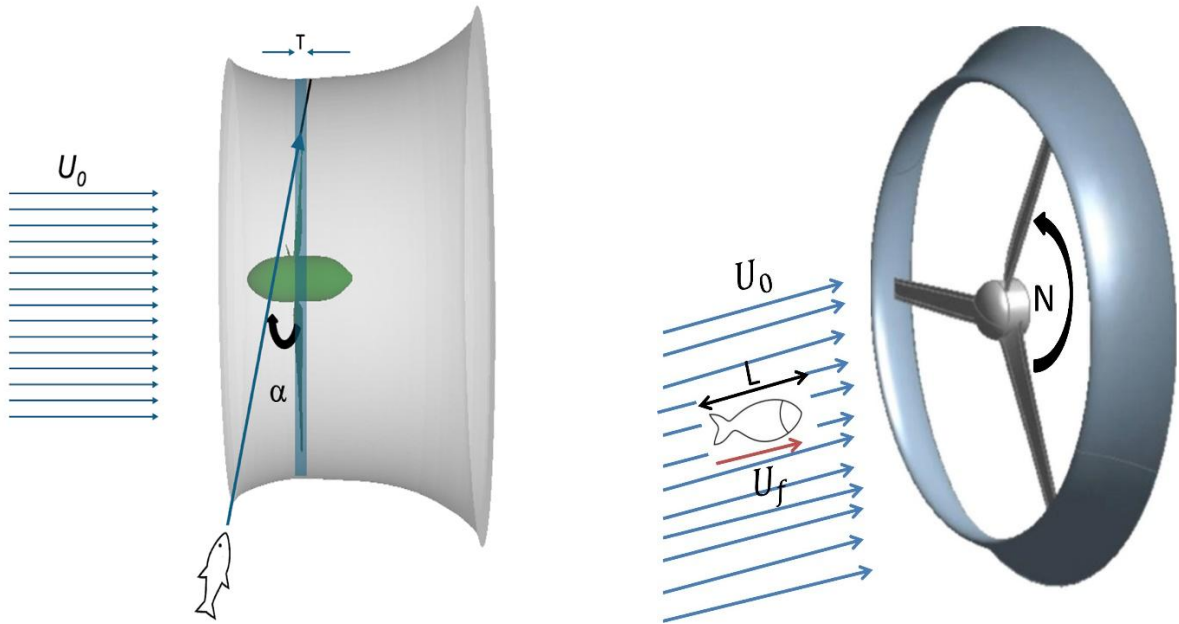

**Fig. 2** Schematic diagram for fish approaching the turbine blades a) Front View b) Side View

The proposed probabilistic model calculates the risk of fish being struck by an individual turbine blade. To assess this strike probability, we developed a Python code to implement the model, incorporating input conditions derived from a typical turbine with a turbine capacity of 5 kW, three blades ($n = 3$), the rotating velocity of the turbine $N = 0.7523$ rotation per second and the thickness of the rotor $T = 0.1$ m.

## 2.2 Fish biological data

To support this evaluation, we gathered the fish species data [40] at the lower Mississippi River site as shown in Table 1.

**Table 1** Fish data with their mean length

| Type | Mean Length $L$(m) |
| --- | --- |
| Blue Catfish | 0.9017 |
| Channel catfish | 0.508 |
| Bluegill | 0.279 |
| Largemouth Bass | 0.5842 |
| Northern Pike | 0.5715 |
| Sauger | 0.4191 |
| Walleye | 0.6858 |
| Shorthead redhorse | 0.491 |
| Shovelnose sturgeon | 0.67 |
| Freshwater drum | 0.3 |
| Smallmouth Bass | 0.375 |
| American Paddlefish | 1.5 |
| Common Carp | 0.74 |
| White Crappie | 0.375 |
| Gizzard shad | 0.45 |

This probabilistic model can support the development of strategies for mitigating collision risks. These strategies may include introducing meshed grids/nets to prevent fish from entering the turbine area and the duct to limit the angle of the fish entrance. The goal is to reduce the environmental impact of HKT installations while maintaining energy production efficiency, thereby aligning technological advancements with global conservation efforts. We will show our analysis in Section 3.

## 2.3 Wake effects of protection grids

Fish protection nets help reduce the probability of strike further, along with our ducted turbine design. However, it is a concern that the net would affect the performance of the turbine. Specifically, the screening net would partially block the flow and reduce the incoming velocity of the turbine [42,43]. This reduction in velocity can be expressed as in Equation 6 where $V_d$ is the downstream velocity due to the blocking effect of the net, $r$ is the Velocity Reduction Factor and $V_u$ is the upstream flow velocity before net using [16].

$$V_d = rV_u \qquad (6)$$

The fact that net structures retard the flow through the fish net has been observed by experimental and CFD studies [17,42,43,44,45]. These characteristics include the Net Solidity ($S_n$), Net Twine Diameter ($d_w$), Mesh Bar Length ($l_w$), and Upstream Flow Velocity ($V_u$).

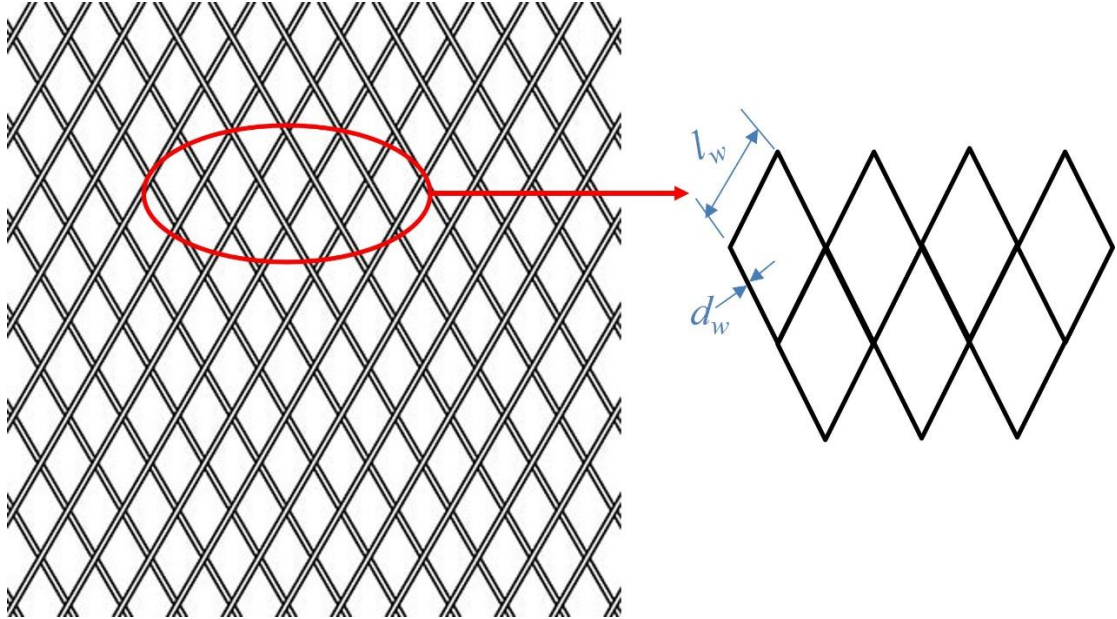

**Fig. 3.** Schematic diagram of the fish net

The flow behavior study from the environmental CFD analysis showed high velocity and high turbulent flow at the targeted site. To limit the blocking effects from the protection grid, we are considering using a steel protection grid to minimize the strike probability at this site. For steel, the net solidity varies between 0.19 and 0.790 as discussed in [16] ($0.19 < S_n < 0.790$). The Velocity Reduction Factor (r) is calculated using the model discussed in [44] and is expressed as shown in Equation 7, which also considers the bio-fouling effect of these netting grids.

$$r = -0.3517 S_n^2 + 0.0256 S_n + 0.985 \qquad (7)$$

Based on the listed equations, a contour plot has been plotted using Solidity ($S_n$) and Upstream velocity ($V_u$) as the axis with Downstream velocity ($V_d$) as the function value as shown in Fig.4. It shows that as the net Solidity ($S_n$) increases, the downstream net wake velocity ($V_d$) reduces with increasing incoming Upstream Velocity ($V_u$).

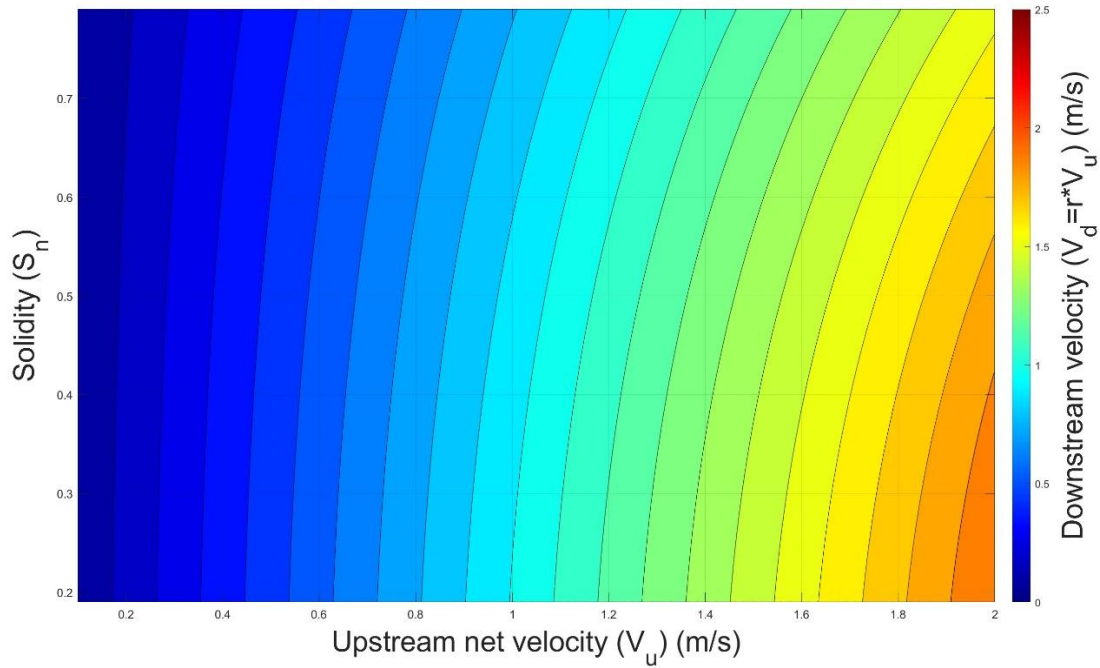

**Fig. 4.** Contour plot highlighting the dependence of Wake velocity on Net Solidity and Upstream velocity

## 3. Results and Discussion

### 3.1 Model Validation

To validate the developed model, we compared the model with lab experiments. The data from [46], which summarizes the mean mortality rate for various species associated with different turbines in hydropower systems, was used. The mean mortality rate for each species associated with the turbine type is summarized as a range using different studies in [46]. The model validation is performed by comparing the model results for the fish species data collected from the lower Mississippi River with the study results corresponding to the same species with the ducted turbine design entry angle. This comparison in Fig.5 showed that our model is well within the range highlighted by the study strike risk results for most of the species. The only exception is for Blue

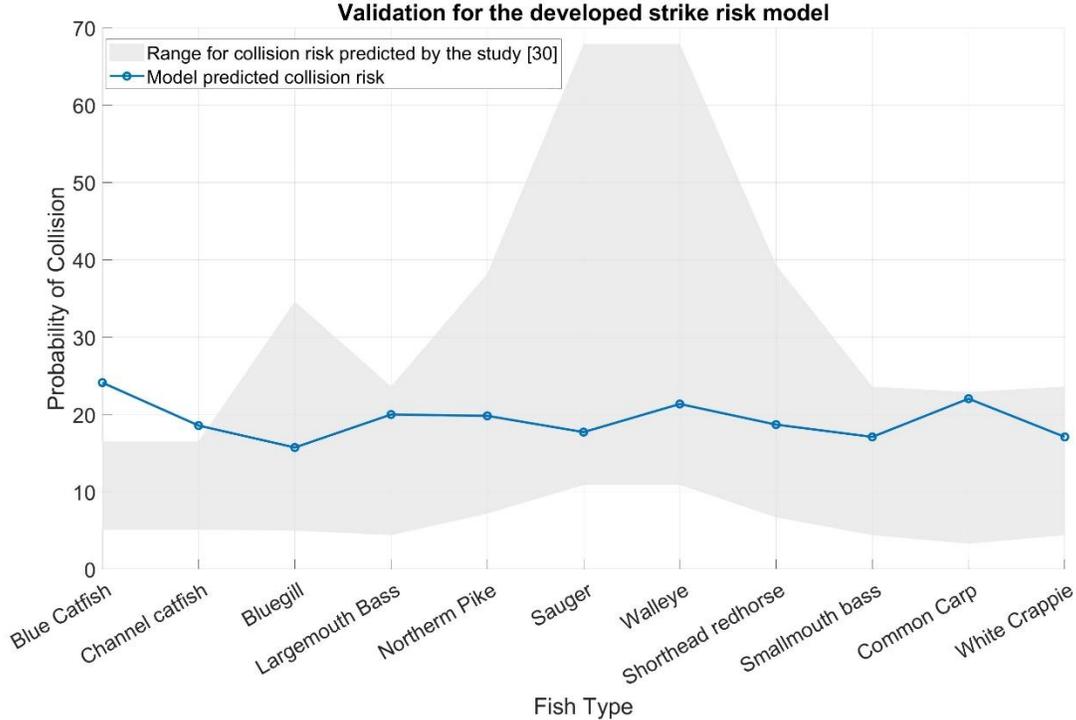

**Fig.5.** Comparison of model collision probability against study predicted range for different fish species

Catfish and Channel Catfish. The possible reason for the mismatch is that Catfish family species exhibit benthic (bottom-dwelling) behavior, with distinct swimming patterns, sensory perception, and reaction strategies that could significantly lower their strike risk than the assumed passive passing process.

## 3.2 Collision risk

Based on the varying fish length, the strike probability is plotted as a contour as shown in Fig.6. This figure shows that the strike risk decreases at the river bottom ($U_0 = 0$) with the entering angle $\alpha$. The lowest strike risk was found for fish with a width of 4 cm. The increased collision risk for larger fish is consistent with the findings of [32], which identified a higher likelihood of turbine blade strike for larger fish size. At the middle depth ($U_0 = 1$ m/s), the strike risk increases with α for large fish but decreases with smaller fish. At the water's surface ($U_0 = 1.7$ m/s), the strike risk decreases with greater α and increases with fish width.

In the design of our ducted hydrokinetic turbine, we've incorporated a feature that significantly enhances environmental safety. The turbine's duct limits the angle at which fish can enter the turbine rotor, ensuring that the entering angle, denoted as α, is always greater than 13.5◦. This strategic limitation of the entry angle effectively reduces the likelihood of fish collisions with the rotor for low river velocity. For example, when $U_0 = 0$, the maximum probability of strike will be reduced by 42%. Hence, we suggest using a meshed net at the inlet to bring down the strike probability further if a turbine is planned to be installed with high flow velocity. A protection grid has been strategically designed to be installed at the entrance of our hydrokinetic turbine, adding an extra layer of safety for aquatic life. Fig.7. also shows the effect is more significant for high river flow if the protection grid can screen large fish. The graph demonstrates how the presence of the grid effectively reduces the likelihood of fish collisions. By physically blocking larger fish from entering the turbine and diverting them away from the rotor blades, the grid ensures that only smaller, more agile fish—which are better able to navigate through the grid—can pass closer to the turbine. This design mitigates environmental impact and aligns with our commitment to sustainable and responsible energy generation.

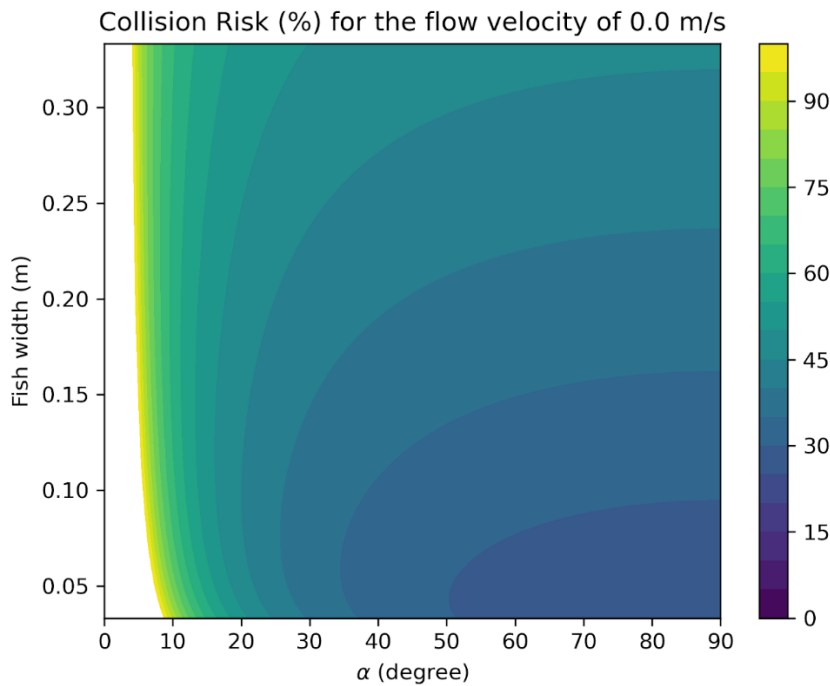

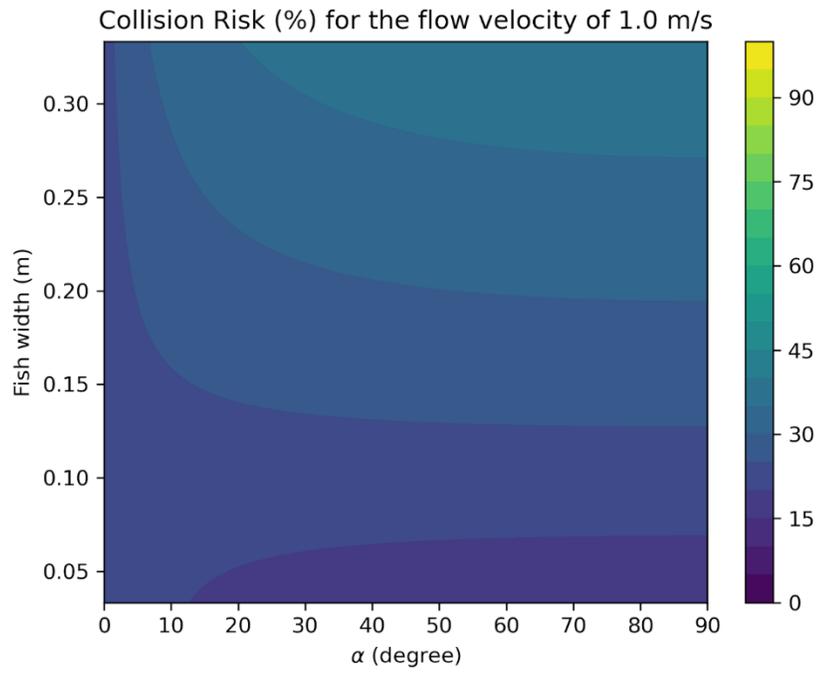

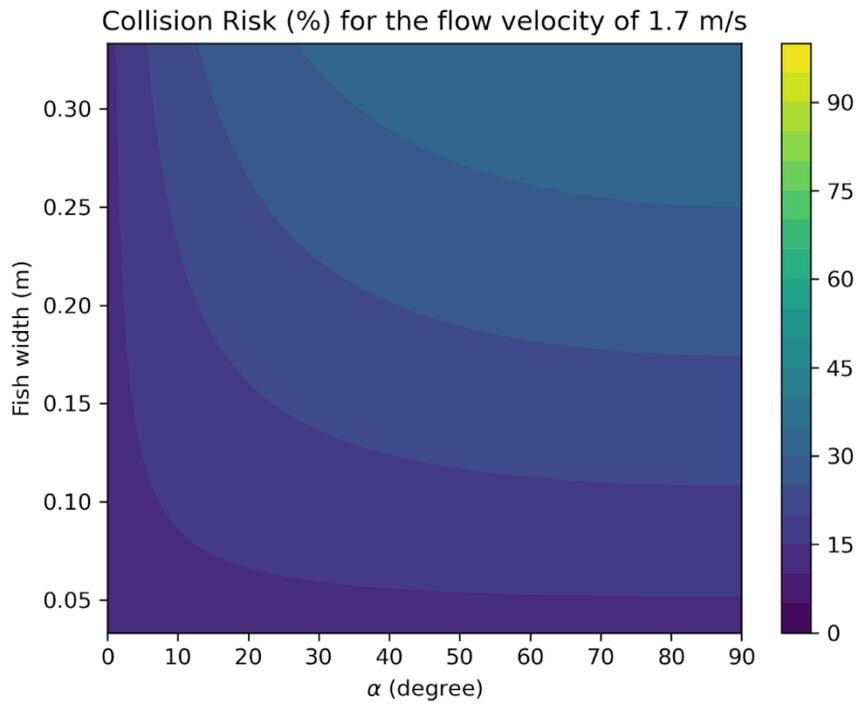

**Fig.6.** Collision risk for the a) bottom, b) middle, and c) surface of the river

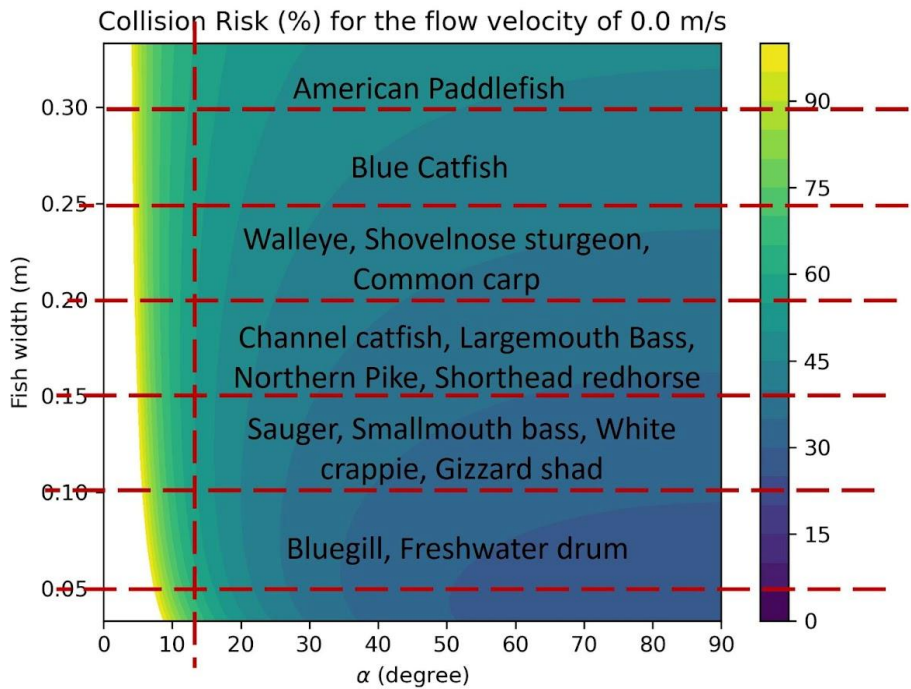

**Fig. 7.** Protection grid sizing based on the fish list at the S3 site.

Despite the valuable insights gained from this study, we recognize a couple of limitations that may prevent the study to scale from site to site or be applied widely. First, the study assume that passive fish swimming mechanism and sensing and navigation strategies were not explicitly incorporated into the developed model for the collision risk assessment. In other words, this model primarily focused on physical parameters, such as turbine blade speed and flow velocity, but did not account for more complex, dynamic behaviors of fish, such as their ability to avoid obstacles in their path, changes in movement patterns due to environmental stimuli, or varying response to turbine presence over time. Nevertheless, this study developed a reliable and convenient model that can provide the first order estimate of fishing collision. Involving detailed turbine designs makes this study enables environmental co-design, i.e. designing environmental risk mitigation scheme from the beginning of the mechanical design.

Second, this study simplified the shape of fish species for convenience of model development based on the fish species at the targeted site, which might limit the generalization of the findings to other habitats with different fish species and environmental conditions.

In addition, the open field observation data is highly limited. Future research should expand the geographical scope, validate the model with field data, and incorporate fish behavior to improve the accuracy and reliability of the collision risk assessment required for environmental permittance of this zero-head technology.

# 4. Conclusion

In this study, fish collision risk was evaluated for the ducted and unducted turbine design using the developed probabilistic model with flow condition derived from the gauge data for the river flow in Baton Rouge, LA. A detailed fish-blade collision model is developed which involves important dimensions of the turbine blades and accessories for improved estimates of collision risk. This study reveals that large fish always have a higher collision risk. When the background flow velocity is low, fish entering with a small angle are subject to higher collision risk, but at high flow velocity, this trend is the opposite. This study further revealed the ducted turbine design limits the angle at which fish can enter the turbine rotor (α >13.5 degrees)- and reduces the likelihood of fish collisions with the rotor for low river velocity. In addition, the protection grid effectively exclude species from the likelihood of fish collisions in addition to our ducted RAFT turbine design, but it compromises the efficiency of the turbine performance.

This study also demonstrates that more experimental and observational data particularly for hydrokinetic turbines in open field is needed for precise validation of this model over the depth of the channel. Future work should focus on the addition of navigation and sensing mechanisms to improve the accuracy of this model.

| Nomenclature | |
|---|---|
| $A_f$ | Cross-sectional Area of fish |
| $C_d$ | Drag coefficient |
| $C_s$ | Swimming force coefficient of fish |
| $d_w$ | Net Twine Diameter |
| $F_d$ | Drag Force |
| $F_s$ | Swimming force by the fish |
| $g$ | Gravity |
| $k$ | Collision coefficient |
| $l_w$ | Mesh Bar Length |
| $L$ | Mean Fish Length |
| $L_x$ | Normalized Fish Length |
| $M$ | Normalized Rotation Speed |
| $n$ | Number of blades |
| $N$ | Rotational speed of the turbine |
| $P_{str}$ | Probability of strike |
| $r$ | Velocity Reduction Factor |
| $S_n$ | Net Solidity |
| $T$ | Rotor thickness |
| $U_0$ | Velocity of the river |
| $U_f$ | Relative velocity of the fish |
| *V<sub>axial</sub>* | Axial water velocity at turbine blades |
| $V_d$ | Downstream flow Velocity |
| $V_u$ | Upstream flow Velocity |

**Greek Symbols**

| | |
|---|---|
| $\alpha$ | Entry angle/Angle of fish relative to the turbine rotor |
| $\rho$ | Density of water |
| $\rho_f$ | Density of fish |

## 5. Acknowledgement

The authors would like to acknowledge the support by the U.S. Department of Energy (DOE), (ARPA-E), (SHARKS).